# Distributed Computation Offloading of an application from mobile/IoT device to cloud

Arsalan Liaqat
*Dept. Of Computer Science*
FAST-NUCES
Lahore,Pakistan.
arsalan.liaqat@lhr.nu.edu.pk

Dr.Saqib Illyas
*Dept. Of Computer Science*
FAST-NUCES
Lahore,Pakistan.
Saqib.ilyas@nu.edu.pk

Ghazala Mushtaq
*Dept. Of Electrical & Computer Engg.*
COMSATS-Lahore
Lahore,Pakistan.
Ghazalamushtaq@cuilahore.edu.pk

*Abstract—* **In Covid-19 pandemic, the number of users connecting to the Internet using mobile devices increased. People are doing there every task using mobile phones [16]. These devices are battery-powered and have limited computation capabilities. Their computational capabilities can be enhanced by computation offloading in which required computation is to be done on a third-party server on a cloud instead of the device itself. The cloud offers virtually infinite computation and storage. We proposed that by exploiting parallelism within an application call hierarchy we can decrease the execution time of off-loadable parts and minimize data resend in case of VM crash. We determine function call paths that are independent of each other within an application and schedule each of them on separate VMs in a distributed way. Wherever such independent paths merge, we collapse to a single VM and whenever the paths diverge again, we schedule multiple VMs. If any single VM fails another copy will be created. However, only the code and data associated with the crashed VM needs to be re-transmitted from the client device. In the case of face reorganization application and montage application we decrease execution time to 27.5% and 43.43% respectively. Whereas the data resend in case if any of both VMs crash will be the portion of the application that had been offloaded to respective VM at depending upon the level of parallelism they have which save mobile battery in case of Resend. We will also discuss the energy consumption effect of using multiple Vms for a job VS single Vm for the same job.**

## I. Introduction

With the increase of IP-based devices in the IT world new research horizons are open for research. The term of IoT and Smart Mobile Devices become famous and force researchers to think upon to make it possible for low computation and limited resources devices to run high computation and resource-demanding applications. The term mobile cloud computing was introduced to solve the said problem. Mobile Cloud offers infinite large resources to mobile devices and allows them to run these resources demanding applications on the cloud instead of these devices. The data or code is offloaded to the cloud for the computational tasks. But it is not the case like we do from our desktop Pc or other high resource and power full machines of nowadays which have high storage and network bandwidth available. There are many aspects that we need to consider before adopting the Mobile Cloud for computation offloading. Mobile Cloud basically do integration of mobile environment with cloud technology and remove hurdles that resist the performance of Mobile devices by offering them resources need to do some task. A mobile device create it link with cloud using wireless medium through base station.

Cloud can be Public Cloud, Private Cloud, Community cloud or hybrid cloud. Some famous cloud service providers are Amazon, Microsoft AZUR, etc. According to statistics given in [12] the mobile devices in 2018 are 4.57 billion while the projected devices in 2020 will be 4.78 billion. whereas in 2007 desktop pc were 75% of internet devices and in 2018 it decreases to 10% only [13]. The mobile industry is capturing the market. The increase in SMD from 2017 to 2018 is 0.1 billion. The world population is almost 7.593 billion and 60% of it is SMD users. Whereas if we talk about all the IP devices connected with the internet including SMD, wearable, IoT devices, etc. the Strategy analysts [14] said that in 2020 all IP devices will reach 30 billion. But it can move up to 50 billion depending upon the market trend.

The motivation behind mobile cloud computing was that IOT and SMD devices have batteries, finite size of memory, and low computation power. Normally the cloud are placed in less populated areas and have enough power supplies and back up to accomplish all the tasks that are assigned to a cloud uninterruptedly and there are a vast variety of studies that debate on how to make the data center consistent, always on and energy-efficient but here that is out of our context of the topic.

Offloading is a challenging aspect of Mobile cloud. Offloading mean moving some data to cloud but it's not an easy task. The decision to offload an application or not, and even if to offload then do full offloading or partial offloading? Is very critical. The factor that contribute towards that decision are Network condition, available bandwidth, data loss ratio ,size off data to offload, execution time of offloaded component on cloud and mobile, resource requirement for doing the task, dependencies between offloaded part and un offloaded parts and user mobility. These devices are connected to base station using wireless medium. Each base station manages the traffic of its region and forward the data packets where they are intended to be forward. In wireless connection the bandwidth is shared and collision can occur easily. Although Carrier-sense multiple access with collision avoidance techniques are used in wireless communication but the still did not guarantee about collision happening. What if we send large data over wireless channel and collision occur then it means we need to send data again and by that practice extra battery of mobile device consumed. Likewise, if collision is avoided but congestion happen in the network it may also result in the data loss. If we are sending same data because of any reason which cause data loss at any point between mobile and cloud that will result in different way first it result in more battery use from mobile and secondly it increase execution time of the application because the task will restart from starting point or from point whose finger prints were saved before execution stop because of data loss .

The increase in execution time depends upon the amount of data that need to be resent and network conditions. Therefore, the decision and methodology of offloading are very important. There are Different offloading frameworks are proposed by researchers all have their pros and cons. All these frame work deal in sending data to a cloud a single VM. In our framework, we are doing offloading on multiple virtual machines depending upon the dependencies between off-loadable data.

## II. OVERVIEW OF OFFLOADING AND MOBILE CLOUD

Offloading is a critical decision in mobile cloud computing. To do offloading different partition algorithms are advised that consider different parameters and advice about offloading whether is should be done or not and in what way. Partition algorithms are the basic decision-maker. In [1] the author gave a survey about mobile cloud computing and discuss its architecture and different factors attached with MCC. In basic architecture, the mobile devices are attached with their base station with wireless medium and by backhaul network, these base stations are attached with external world network these base stations act as a gateway for mobile devices and make it possible for these devices to communicate on WAN or LAN by doing due transitions. Then the link goes to the cloud where many servers are available.

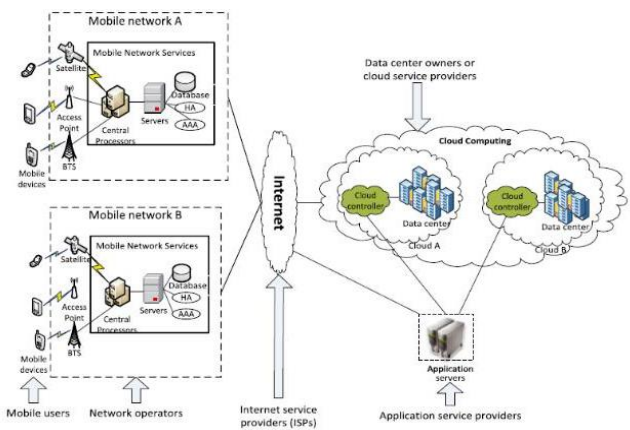

Fig.1  Mobile cloud architecture

Figure.1 shows the basic architecture of the mobile cloud. We borrow them and pay in return for cloud services that we use for computation or storage. The cloud can be distributed at different geographical locations or may be contained within a small building it purely depends upon the type of cloud and scale of the cloud. The cloud is controlled and managed by the cloud controller of the respective cloud. Cloud provides three types of services to its user which are platform as a service (paas), infrastructure as a service (iaas), and software as a service (saas).what type of service we need to depend upon the type of control we need on a virtual machine in the cloud. Offloading can be full or partial [2]. Programs contain different parts some of them are off loadable and others are un-off-loadable. They may or may not have dependencies among them [10]. Whereas offloading is not the best decision always because for offloading we need to do some extra work for transmitting the data over the network to the cloud and also pay the cost for borrowing cloud resources.

There are two costs associated with offloading one is computation cost the other is communication cost. By computation cost, we mean cost vs benefit comparison whereas by communication cost we mean time taken by data to travel over the network from device to cloud and mobile energy used in transmitting data. In offloading First the data is transmitted to the cloud then the cloud performs computation on the data and sends the result back to the mobile device. Whereas the data send on the cloud from mobile is larger in size while the data or results send back by cloud to mobile are smaller in size. We generally say there are three-step in offloading in fist we initialize the application in the second we offload the computation and third execute an application on the cloud There are frameworks that suggest VM-based computational offloading [3]. There are different frameworks are suggested for computation offloading. The concept of computation offloading is not new it was proposed in 1970 but become popular in industry after the invention of SMD and wireless communication invention. Before computation offloading few aspects are considered [11]

1. User preferences that the user want the data to be confidential or not?
2. Network specifications like WiFi,3G ,4G etc.
3. Device Specification
4. Application specifications like its resource requirement
5. Server Specification on which the application is offloaded

After considering the above mention factor the decision of offloading is taken. Offloading can be data offloading or computation offloading. In data offloading the storage capabilities of the cloud are used like iCloud etc. whereas in computational offloading the computation capabilities of the cloud are used. Figure 2 describes the relation of both offloading. Offloading has different cons associated with it, but we are addressing two issues.

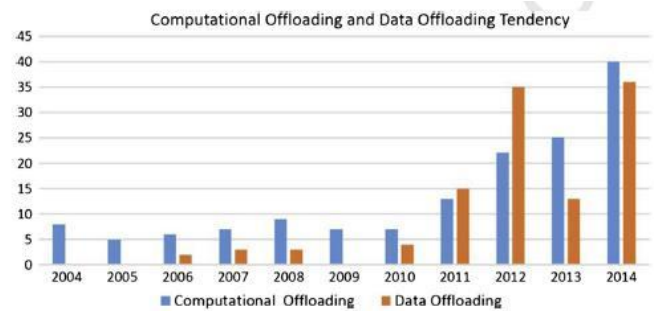

Fig.2  Computational offloading and Data Offloading Tendency

One is to minimize execution time on the cloud and the second is to minimize data resend in case of data loss because of Vm/server went down or network condition cause data loss.

Com during offloading. Till now no framework for offloading applied parallel computation technique on offloading for one application. We are proposing a parallel offloading technique that decreases our computation time and if VM crashes then we need to resend partial data rather than the total offloaded data.

## III. PREVIOUS OFFLOADING ALGORITHM

Now we will discuss some previously proposed offloading techniques.

### A. Kumar And Lu Partition Algorithm

They use two parameters to take the decision of offloading one is computation time second is communication time [8]. The partition decision is taken on the trade-off between the above-mentioned two parameters. Their results show offloading is beneficial if we need low communication for high computation. They advise an equation as given below. The parameters are described below C=computation need M=speed of mobile processor S=speed of server processor D=data to send B=bandwidth P=where P is the

$$P_c \times \frac{C}{M} - P_i \times \frac{C}{S} - P_{tr} \times \frac{D}{B}. \quad (1)$$

Suppose the server is $F$ times faster—that is, $S = F \times M$. We can rewrite the formula as

$$\frac{C}{M} \times \left(P_c - \frac{P_i}{F}\right) - P_{tr} \times \frac{D}{B}. \quad (2)$$

power.

### B. Hengwang And Zhiyuan Li Partition Algorith

They proposed a scheme by translating the application into a serve-client model structure and dividing the code into parts or sub-problem where each part or sub-problem is a task [5]. Some code run on the client-side (mobile) and some run on the server side(cloud). The server-side code is offloaded as part of the application. They also consider communication and computation cost in their scheme but the difference between this approach and the previous one is that they also use the cost of run-time bookkeeping which keep record or distribution and is very important for distrusted execution. This algorithm is distributed so there can appear time gaps between the execution of tasks on the client-side and server-side.

### C. Adaptive Comutaion Offloading

This algorithm uses online statistics of network conditions and RRT time. These are used to calculate break-even time which is the minimum time for computation [9]. When we run any task on client or mobile we consider it as a timeout timer if in that time the client or mobile fails to complete the task then we move the task to the cloud. The drawback of this algorithm is that when an application is offloaded to the server it already consumes beak even time and energy associated and then sent to the server where it again starts from the initial point of the task. Execution time does not need to be known prior because we are not using them in the decision of offloading as we are using break-even time.

### D. Xiamoning Fus Partition Algorithm

This approach considers different parameters to decide offloading is to be done or not. This algorithm dependencies between different methods of an algorithm are notified and are represented in a call link graph using some reverse engineering tool.it uses data upload, return time, execution time on the cloud, and execution time on mobile. Using these parameter partition algorithms decide which methods of the algorithm can be offloaded and which part will be executed on a mobile device [4].

$$U_i = \begin{cases} \max_{1 \le k \le K} \left\{ \sum_{j=1}^{k} (M_j - C_j) - I_1 - R_k \right\}, & \text{if } i = 1 \\ U_{i-1} - (M_{i-1} - C_{i-1} - I_{i-1}) - I_i, & \text{otherwise} \end{cases}$$

## IV. DISTRIBUTED OFFLOADING

It is described in [4] that applications are composed of different methods. These methods have a call hierarchy in them to execute. These methods are furthermore classified into off-loadable methods and un-off-loadable methods. When we want to perform some computational task on the cloud we offload respective methods on the cloud. Off loadable methods are those that have less communication cost and high computation benefits whereas un-off-loadable parts have more communication cost and fewer computation benefits. The call hierarchy of methods can be expressed by making a call link [4] graph.

it is observed that some application has a tree-like structure within and have a sequence of methods execution. When they are converted into a call link graph based on call hierarchy. it is been observed applications have different chains of methods calls and these chains are independent of each other. These chains can be fully/partially off loadable or fully or partially un off loadable. We can offload these independent chains in parallel mode onto different VMs. we can minimize the execution time and data resend in case of data loss occurs because of any reason. Data loss can occur because of network conditions, Vm crash, server failure etc.

We offloaded independent computation chains in a distributed way on different VMs. We use a simulator CloudSim for the experiment on which we created different numbers of VM of the same specification for the same application to observe behaviors. Our VM have 1024Mb RAM, 10000 MIPS CPU with Two Processing Element (PE). Whereas the system on which we run the simulator has 4GB RAM, Core m processor. Core m is a new generation of Intel which is designed by keeping mobile technology in mind and these processors dissipate less energy. We measure executing time, data resending rate, and power consumption using parallel offloading. For that, we created numbers of cloudlets and execute them on different VMs with the same configuration.

### A. Time Saving in Case of Distributed Offloading

The time saving by using distributed offloading depends upon two factors
1. number of independent chains within an application
2. number of methods we can offload

If we have ten offload-able parts but they all belong to one chain, then there will be no time saving. We can save execution time only if we have multiple independent off loadable chains. The task execution time will be the maximum time taken by the chain among all the chains. The execution time depends upon the time taken by the chain but not on the number of methods it has. For example, if we have two chains chain A have 5 methods whereas chain B has 2 methods, we cannot say that chain A will take more time because it has more method. Maybe the method belonging to chain A is less computation expensive than chain B methods which in result take more execution time than chain A methods.

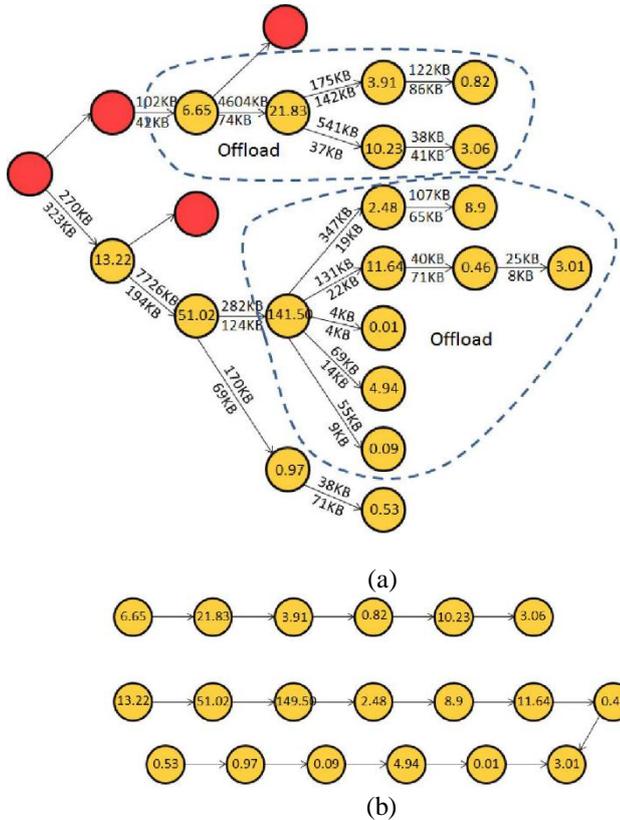

(a)

(b)

Fig.3    The call link graph of face recognition application

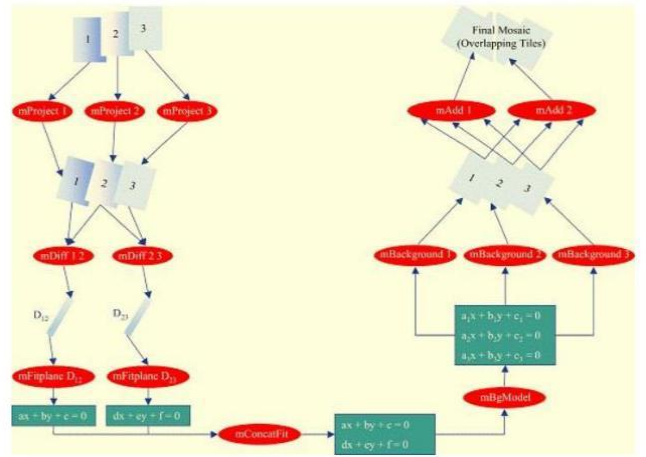

Fig.4    The flow chart of montage application

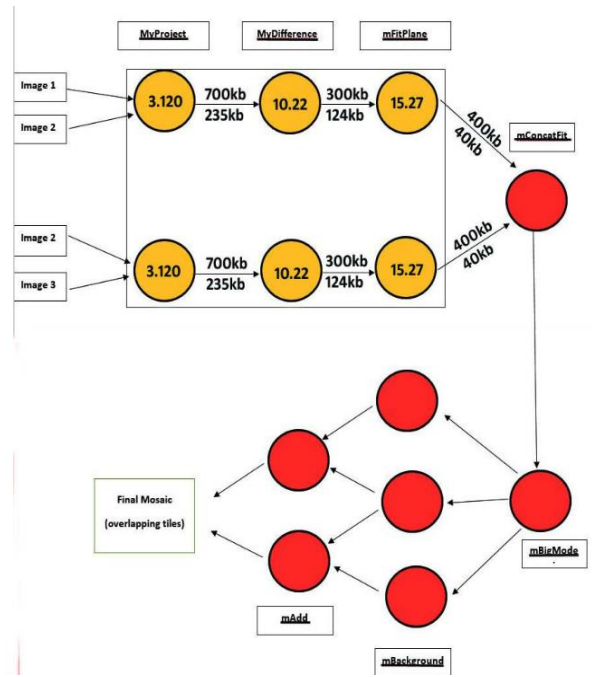

Fig.5    The Call link graph of montage application

We used face recognition and montage application to prove our idea and result. The call link graph of face reorganization application is given in Fig .3. Each circle indicates a method and links show their call dependencies [4]. Red circles are un- off- loadable methods and yellow circles are off-loadable methods. Figure 3. (b) Shows two different call links which are independent of each other. The naive approach is to offload the whole graph sequentially whereas in distributed offloading we will offload these independent call graphs onto different VMs. Table 1 shows the result that using distributed offloading we minimize the execution time by 27.5% which is a quite nice improvement.

Table1: execution timings for Face reorganization Application

| Offloading Strategy | Execution Time (Sec) |
|---|---|
| Sequential/Non-parallel | 1.2 |
| Distributed/parallel | 0.87 |

Table2: Execution timings for Montage Application

| Offloading Strategy | Execution Time (Sec) |
|---|---|
| Sequential/Non-parallel | 0.76 |
| Distributed/parallel | 0.43 |

In Case of Montage application after studying its graph, which is shown in figure 4 we can see some of its parts can be offloaded in parallel.in figure 5 the dotted area is the part of application that we can offload on cloud which result in less execution time for the application.

Data related its execution times on cloud and mobile device are not available, so we hypothetically assume some data for it and perform experiment on that which showed the execution time is minimized 43.43% in case of distributed execution in Table 2.

## B. Data Resend If VM Crash in Case Of Distributed Offloading

The offloading involves sending of data to cloud. This data can be small as to some Kbs or can be large up to hundreds of Mbs depending upon the type of task need to be executed. Sending a large amount of data from Mobile/Iot device to cloud need battery and effort of these lower computation and resource device. What if data is lost? They need to resend data with ask for extra work for these devices in resending data.

If we use distributed offloading, we can relax these devices in these scenarios of data loss up to some extent and save there battery power. Let say we have two independent chain and we offload them on cloud. Let say each chain is of 100 Mbs so collectively we are offloading 200Mbs.If Vm destroy and we need to send a copy of it on cloud ,in conventional approach we will resend 200mbs but if we are using the advised distributed approach and offloaded the date on two Vm ,each for independent chain, suppose Vm A went down so we will resend only data associated with Vm A which in this example is 100Mbs.

How much data we need to resend in case of data loss because of Vm crash or network condition depends upon level of parallelism present in call link graph of respective application and the data associated with the respective Vm.

The higher the parallelism we have between off-loadable methods the lesser will be the data resend as shown in Table3 assuming at any time one Vm can Crash because Vm are distributed. It's been clear that the higher the level to parallelism we have between different independent call link the minimum will be the data re send in case of VM crash or any other reason which results in data loss.

Table3: Data resend with respect to level of parallelism

| Total % Offloadable | %Offloaded to Vm1 | %Offloaded to Vm2 | Max data Resend% |
|---|---|---|---|
| 70% | 70% | 0% | 70% |
| 70% | 60% | 10% | 60% |
| 70% | 50% | 20% | 50% |
| 70% | 35% | 35% | 35% |

## C. Energy consumption comparison

Using multiple Vm instead of single Vm proves as a better option in term of reducing execution time and data resend. In current time all clouds and data center are movie toward green Cloud, which mean they are becoming more efficient in term of energy consumption.to measure our strategy in this paradigm we create different cloud lets of same type and run them in both scenarios. First, we run them on single Vm and measure the energy consumption at the cloud then we distribute them into two parts and run them on two separate Vms.we tested 10 cloudlets on two Vm and the results show the energy consumption is similar but this may increase slightly if we over load the real hardware who is running these Vms.The Results are displayed in Figure 6.

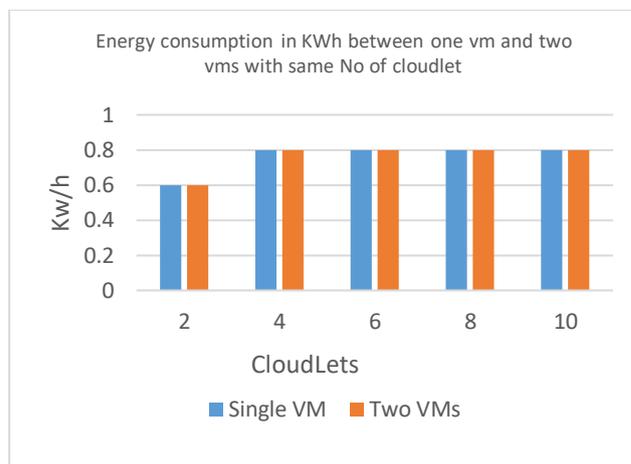

Fig.6 Energy consumption in KWh between one VM and two VMs with same No. of cloudlet

## V. CONCLUSION

The results show that if applications have multiple independent off loadable methods/jobs then it is favorable to execute them on different VM because that results in execution time saving, fewer data resend ratio in case of VM went down by which our battery time extends. The Cloud energy consumption remains unaffected because the energy factor is mainly associated with real hardware. Using multiple Vms may cause some extra cost to pay but application speed and stability will increase. Our proposed strategy reduced the execution time of face reorganization application to 27.5% and montage application by 43.43%.